\documentclass[10pt]{article}
\usepackage[parfill]{parskip}
\usepackage{geometry}
\usepackage{enumerate}
\usepackage{graphicx}
\usepackage{mathtools}
\usepackage{url}
\usepackage{authblk}
\usepackage{hyperref}
\begin{document}

\title{A PLS blockchain for IoT applications: protocols and architecture}
\author{Alex Shafarenko\footnote{Correspondence address: University of Hertfordshire, U.K.; 
email: \href{a.shafarenko@herts.ac.uk}{a.shafarenko@herts.ac.uk}}}
\affil{University of Hertfordshire, AL10 9AB, UK; Ada Finex Ltd}
\maketitle

\def\things{{\it things}}
\def\thing{{\it thing}}

\begin{abstract}
This paper proposes an architecture and a protocol suite for a permissioned blockchain for a local IoT network. The architecture is based on a sealed Sequencer and a Fog Server running (post-quantum) Guy Fawkes protocols. The blocks of the blockchain are stored in networked Content Addressable Storage alongside any user data and validity proofs. We maintain that a typical IoT device can, despite its resource limitations, use our blockchain protocols directly, without a trusted intermediary. This includes posting and monitoring transactions as well as off-chain (post-quantum) emergency communications without an explicit public key.  

{\bf keywords:} blockchain, Guy Fawkes protocol, post-quantum, HORS-OTS, LoRa, concurrent transmission
\end{abstract}

\section{Introduction\label{intro}}

This article presents a Block Chain construction based on the well-known Guy Fawkes Protocol (GFP) \cite{GFP} for digital signature, which we extend and bring to bear on Block Chain (BC) technology intended for a swarm of low-power devices (IoT \things). The primary purpose of this blockchain is to support an immutable distributed ledger that ensures the authenticity and sequencing of user records posted on it. Financial transactions for IoT are {\bf not} our intention, but they should be compatible with our approach.

We set ourselves the following design constraints on behalf of the participating \things{}:
\begin{enumerate}
\item Post Quantum restriction, in particular no public key crypto
\item low power, low energy. Notice that the avoidance of public key crypto is synergetic with this constraint 
\item low local storage. A \thing{} may have a flash card embedded in it, but the use of the flash card eats into the energy budget
\item \label{local-comms} local communications with low bandwidth and short messages. The security protocol should be conducted via UHF broadcasts over the target area. An effective adversary should have to radiate significant power (reliably over the legal limit) and expose itself to triangulation
\item it must be possible for all \things{} to authenticate the ledger under realistic assumptions without relying on a trusted intermediary 
\end{enumerate}

Those are the main objectives. We believe the main security risk with our scheme to be the Denial of Service (DoS) attack. Due to the use of local communications (constraint \ref{local-comms} above), DoS attacks only need to be impeded but not totally suppressed. The latter is impossible due to the possibility of an attacker's physically jamming the communication infrastructure. 

Section \ref{idea} introduces the protocol idea.  Section \ref{arch} defines the system architecture.  The protocol for posting content on the blockchain is presented and discussed in Section \ref{sec:SLVP}. In the subsequent section, Section \ref{enrol}, we show how a user can be enrolled on the chain at a point other than the beginning.  The next section puts forward a solution for emergency communications, when the originator cannot wait for the next block of the blockchain to be published.  Section \ref{related-work} cites some related work and finally there are conclusions.

\section{Protocol Idea\label{idea}}

The Guy Fawkes Protocol family was first proposed by a Cambridge group \cite{GFP} 10 years ahead of Nakamoto. The original formulation is very clear and can be practically useful, but it was not specifically intended as a blockchain protocol. We will now introduce a similar construction optimised for this purpose. 

\paragraph{Goal.} Assume that a single transmitter is to broadcast a stream of public, authenticated messages to an unspecified number of receivers. The following conditions must be satisfied: 
\begin{enumerate}
\item It should be possible (ideally at low cost) for each receiver to prove, without trusting any intermediary, that the message was sent by the transmitter.
\item It should be cryptographically hard for an attacker to modify any message {\bf or} to change the order of messages without the receivers noticing
\item The broadcaster should be able to send an unlimited number of messages without weakening the security of the previous two constraints.
\end{enumerate}

(We will also observe the Post Quantum restriction here: no DH, no public key crypto.) 

\paragraph{Threat model.} 

\begin{enumerate}[i]
\item An attacker can force one or more receivers to receive the attacker's arbitrary message instead of the one being transmitted by the transmitter, or prevent a receiver from receiving the message at all. 
\item However, the attacker cannot thus disrupt {\em all} receivers and it cannot make {\em any} receiver conclude that the broadcast did not take place. The former can be achieved by delivering at least one copy of the broadcast message by an alternative physical channel, and the latter by broadcasting messages on a public wall-clock schedule.
\end{enumerate}

Point (ii) makes it possible for a receiver that the protocol determines has received an invalid message to solicit unauthenticated candidate messages from peer receivers. We will show that using the protocol each receiver will be able to select the genuine message out of a set of candidates. This makes threat (i) a DoS threat rather than a data-integrity one.
   
Now to the protocol. It uses a standard cryptographic hash $H(x)$ (e.g. SHA-256) and assumes that it is sufficiently hard to find a pre-image $x$ given its hash $H(x)$. In fact so hard that it is impossible to find $x$ within the time between the signing of two consecutive messages. All values except $M_k$ are binary strings of the same length as $H$. Additionally the protocol uses symmetric encryption ${\bf E}_q(p)$ which encrypts plaintext $p$ under the key $q$ producing a ciphertext, and its dual decryption: ${\bf D}_q({\bf E}_q(p))=p$. This could be any standard cipher, e.g. AES128, suitably adapted to the key and text size using one of the standard methods.

The protocol operates in steps according to the wall-clock time. All receivers and the transmitter synchronise their clocks so that when the transmitter's clock registers a time $t_T$, any receiver's clock $t_R$ is no more than $\epsilon$ away: 
\[ |t_T-t_R|<\epsilon\,.
\] 

The transmitter broadcasts at regular intervals, $t_0,t_0+\tau,t_0+2\tau,\ldots$, where $\tau\gg\epsilon$. Each broadcast consists of three messages of the same length as $H$: $P$, $L$ and $S$; they are a {\em proof}, {\em link} and {\em signature} message, respectively. It is convenient to think of them as being broadcast on three different channels, or in three different time slots, or with a tag that tells the receiver which message it is. The messages are not explicitly indexed, but it is convenient to think of them as being indexed with the Broadcast Interval Number (BIN): BIN 0 corresponds to the interval $[t_0+\epsilon, t_0+\tau-\epsilon]$, BIN 1 to the interval $[t_0+\tau+\epsilon, t_0+2\tau-\epsilon]$, etc. Note that the $P$, $L$ and $S$ broadcasts in the same interval are not mutually ordered.

\begin{figure}
\hskip-0.05in\bgroup
\def\arraystretch{1.3}%
\begin{tabular}{|c|l|l|l|}
\hline
 BIN& Transmit/Receive & Verify & Obtain\\
\hline
\hline
1&$L_1=H(N_{2}) \oplus N_1$               &&   \\
 &$S_1={\bf E}_{N_1} (H(M_1)\oplus H(N_{2}))$ & $P_1$ out of band &~ \\
 &$P_1=H(N_1)$                            &                           &\\

\hline

2&$L_2=H(N_{3}) \oplus N_2$               &                        &  \\
 &$S_2={\bf E}_{N_2} (H(M_2)\oplus H(N_{3}))$ & $H(L_1\oplus P_2)=P_1$ & $H(M_1) = P_2 \oplus {\bf D}_{L_1\oplus P_2}S_1$\\
 &$P_2=H(N_2)$                            &                        & \\

\hline

3&$L_3=H(N_{4}) \oplus N_3$               &                         &  \\
 &$S_3={\bf E}_{N_3} (H(M_3)\oplus H(N_{4}))$ &  $H(L_2\oplus P_3)=P_2$ & $H(M_2) = P_3 \oplus {\bf D}_{L_2\oplus P_3}S_2$\\
 &$P_3=H(N_3)$                            &                         & \\

\hline
...&...&...&...\\

\hline

$k$&$L_k=H(N_{k+1}) \oplus N_k$             &                                &  \\
 &$S_k={\bf E}_{N_k} (H(M_k)\oplus H(N_{k+1}))$ & $H(L_{k-1}\oplus P_k)=P_{k-1}$ &$H(M_{k-1}) = P_k \oplus {\bf D}_{L_{k-1}\oplus P_k}S_{k-1}$\\
 &$P_k=H(N_k)$                              &                                &\\

\hline
\end{tabular}
\egroup
\caption{PLS Protocol\label{fig:gf}}
\end{figure}

Now to the calculations, figure \ref{fig:gf}. In or before the first interval, receivers obtain independent authentication of $P_1$. In each interval $k$ the transmitter creates a fresh random nonce, $N_{k+1}$, and keeps it secret until the end of the next interval $k+1$. One such nonce, $N_1$, is created by the transmitter before launching the protocol. 

The protocol is a two-stage pipeline. At BIN=1, the transmitter sends out the link, signature and proof messages, saves $N_2$ and keeps it secret till the end of the next interval. The receiver receives and saves the received messages. It uses the remaining time in the interval to poll its peers to learn any alternative values of $L$, $S$, and $P$ should they be received (which may be due to signal propagation problems, deliberate jamming or a cyber attack). 

At BIN=2, the receiver receives $P_2$ (possibly more than one candidate value)and verifies that for some candidates $L_1$ and $P_2$, $H(L_1\oplus P_2)=P_1$. It means that these values of $L_1$ and $P_2$ are genuine. Indeed, an attacker wishing to convince the receiver that an alternative link message 
\[\hat{L}_1=H(\hat{N}_2)\oplus N_1\ne L_1\]
is genuine in order to make a forged signature for its own message $\hat{M}_1$ 
\[\hat{S}_1={\bf E}_{N_1}(H(\hat{M}_1)\oplus H(\hat{N}_2))\]
would have to have sent these messages at BIN=1, when only the transmitter knows the value of $N_2$, so the attacker would have to use its own nonce, $\hat{N_2}\ne N_2$, and then force the receiver to receive $\hat{P}_2=H(\hat{N}_2)$. To succeed at that, the attacker must be able to obtain $N_1$ within interval 1, to use it in the $\hat{L}_1$ message. But all that is publicly known about $N_1$ {\em then} is its hash, $H(N_1)=P_1$. 

That is the linchpin of the security of any Guy Fawkes protocol, our version or the classic\cite{GFP} alike. The attacker has to find a pre-image of a public hash value in order to mount a successful attack. The chances of finding a pre-image are slim,  $~2^{-(l-1)}$; for $l=256$ the attacker would have to do around $10^{77}$ hash calculations to find the pre-image on a classical computer. This can be improved upon by quantum computing, reducing the number to $2^{-l/2}\approx 10^{38}$, still reliably unfeasible and certainly good enough for the realm of the IoT.

Leaving the DoS scenario aside, we assume that the receivers succeed. Next the receiver uses the formula in the last column of the table in figure \ref{fig:gf} to calculate $H(M_1)$ on all candidate $S$-messages recorded at BIN=1. All these values are considered equally valid, but only one of them, namely the genuine value of $H(M_1)$, where $M_1$ is a message known to the transmitter at BIN=1, can ever be used. An attacker corrupting message $S_1$ can only achieve denial of service: to fit a message to an arbitrary hash value of it (for example, by extending the message with a non-data-bearing tail) is computationally as difficult as it is to fit $N_1$ to a known $H(N_1)$. 

The reader may wonder about the purpose of encryption. In the original Guy Fawkes protocol only hashes are used, but at least four hash-length items need to be communicated in each signing round, whereas PLS requires only three, a 25\% saving in communication costs. Communication is important for the IoT world, where typically the radio duty cycle of a \thing\ is limited to a fraction of 1\%. Also we argue that the computational cost of a symmetric encryption is several times cheaper than that of a hash for at least some popular IoT platforms. 

Another question is whether the public availability of $H(M_1)$ gives the attacker an alternative method for obtaining $N_1$ at BIN=1: by brute-forcing the encryption key. The answer is that this would require the plaintext (the cipher text is publicly available as $S_1$ at this point), and to obtain the plaintext the attacker needs to know $H(N_2)=P_2$, which is not revealed at BIN=1. To obtain $H(N_2)$ from the public value of $L_1$, the attacker requires the value of $N_1$ which is what the attacker is trying to fit.

At the end of interval 2 the receiver will have received, and collected from peers all alternative versions of, $L_2$ and $S_2$ and is now prepared for the next, third interval, etc. The protocol is run periodically as long as the transmitter stays in commission. Since any secrets have a short lifetime $2\tau$ after which they are published rather than merely not used, there is no accumulation of confidential material at the transmitter site; consequently the transmitter has no security motivated expiration time. 

\section{System architecture\label{arch}}

The PLS protocol described above is well suited to serve as a basis for a blockchain system. The key property that makes it so suitable is provable, time-referenced {\em forward chaining}, the fact that $L$-messages establish a cryptographically protected, unsplittable temporal chain of $\{M_i\}$, with each $M_i$ being defined by its hash $H(M_i)$. With the head of the chain independently authenticated for all actors before the protocol launch, the set of link/proof pairs and the hardness of the hash pre-image problem guarantee that the chain can be validated in isolation by any observer {\em who is present and able to receive messages at broadcast times}. No additional source of trust is required to validate the chain although a trusted third party may well be useful as defence against a DoS attack, bearing in mind that light touch security would be sufficient for that purpose.

\begin{figure}
\begin{center}
\includegraphics[width=0.8\textwidth]{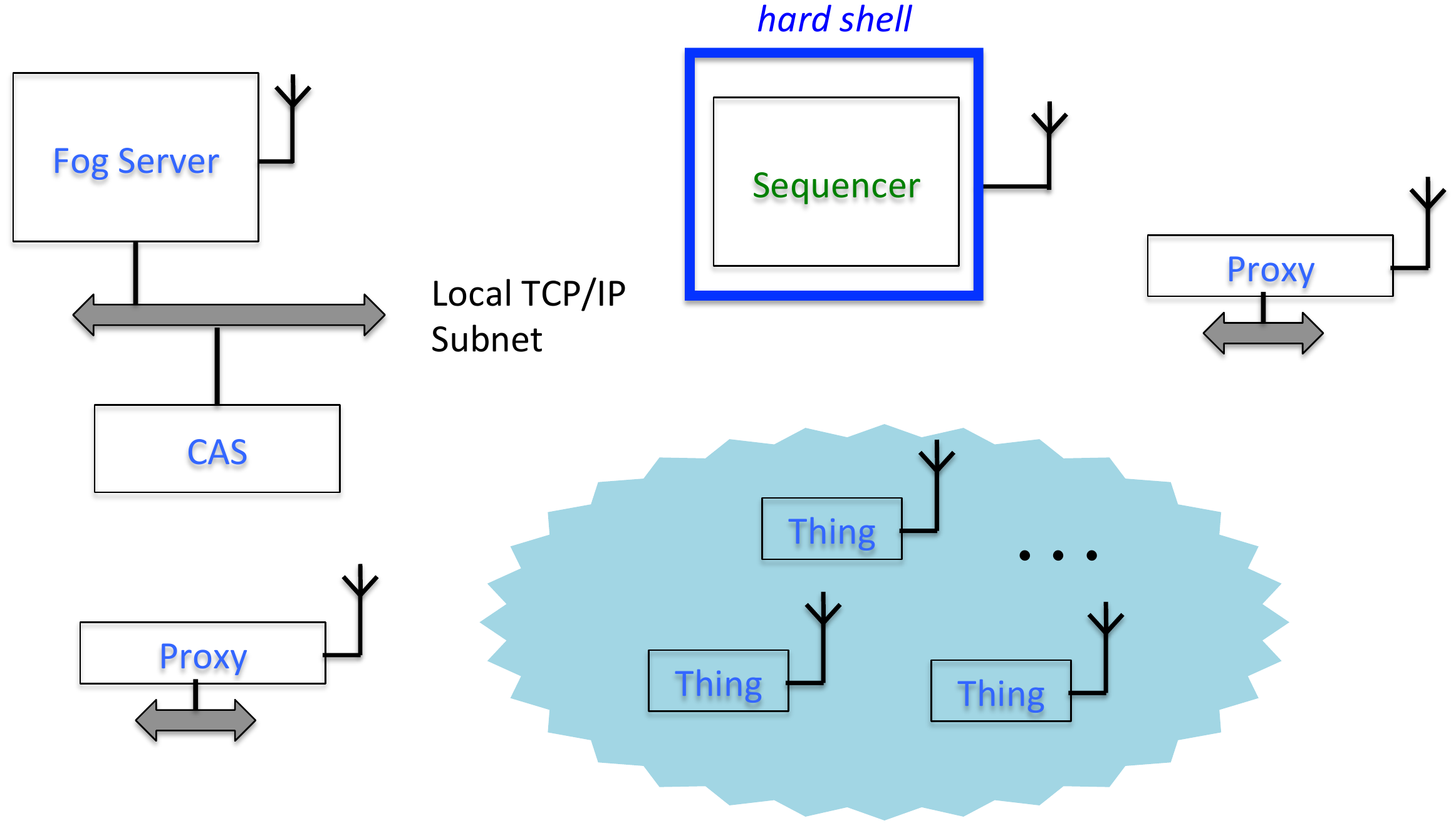}
\end{center}
\caption{Architecture of a PLS-blockchain system\label{fig:arch}}
\end{figure}

We propose an architecture of an IoT system with blockchain services, see fig \ref{fig:arch}. 
\paragraph{Sequencer.} At the core of it is placed a physically secure, firmware based, multi-radio connected blockchain Sequencer. The job of the Sequencer is, as the name suggests, building the sequence of chained blocks. The blocks themselves are prepared for the Sequencer by the Fog Server (see below). The Fog Server communicates with the Sequencer in one way (transmits), over a separate authenticated radio channel (Bluetooth, for example), the details of which are private to them. What differentiates the Sequencer from the Fog Server is the fact that the Sequencer is {\em not} connected to any general purpose networks. Its security role is to run an active PLS protocol on schedule and to keep each nonce $N_k$ confidential until the next broadcast period. The Sequencer's embodiment as a separate air-gapped unit with enhanced physical security serves only one purpose: prevention of a blockchain split, which is effective if and only if the confidentiality of $N_k$ over a short period of time can be assured. Split avoidance is important for defeating DoS attacks, but the validity of the blocks is not in jeopardy due to the second aspect of the Sequencer, namely the fact that it performs a radio broadcast using a wall-clock synchronised, long-distance signal. As a result, if the defences fail and an attacker compromises the Sequencer, the only non-DoS way to profit from this is to have different broadcasts directed to different groups of \things. In this case the inconsistencies will eventually be detected by radio monitoring, but it could be too late, especially if \things\ perform critical work, hence the importance of air-gapping and physical protection for the Sequencer. 

\paragraph{CAS.} For the purposes of sequencing and signing we are going to use the PLS protocol (and a similar SLVP protocol for \things, see section \ref{sec:SLVP}), but what is going to be sequenced and signed are in fact the hashes of the actual blocks and blockchain users' content messages of variable size. To store these items and retrieve them at a user's request, we place a {\bf Content-Addressable Storage} (CAS) unit on a local TCP/IP subnet and organise a private channel (behind the organisation's firewall) between the unit and the Fog Server. CAS  operates in WORM mode: a file is stored once under the hash of its content as its file name, which means that it cannot be changed without the file name changing, that it is easy to verify that it hasn't been changed and that no trust between a user and the CAS is required. Due to the irreversible nature of hashing, it is prohibitively expensive to store a different content under the same hash, so all the content recipient needs to do to ascertain the integrity of the content is check that its hash is correct. 

Every time a new block is formed to go on the block chain, the block content is stored in CAS and the hash of it is used in the protocol. All \things\ have access to CAS via the Server or its Proxy and can retrieve files whenever the content hash is known. 

Note that the hash functions used by CAS and our protocols do not have to be the same. By extending the encryption in protocol $S$-messages to accommodate a longer plaintext and expanding $H(N_{k+1})$ by replication and concatenation, one could use longer hashes for storage than those used in the protocol. It may be profitable to do so given that the second pre-image resistance of a hash in the PLS protocol need only to withstand an attack over $2\tau$, while the resistance of the hash function used to compute CAS file names is potentially required to be much greater  (years if not decades before broken). This is not an argument for a longer hash for CAS (a standard AES256 is quite adequate even against quantum attacks) but perhaps a shorter hash for PLS (128 bits) could be sufficient.

\paragraph{Fog Server.} An IoT system's security cannot be completely decentralised since, ultimately, the network of \things{} is owned by an organisation, and that organisation should have authority to add and remove \things{}, configure them, assign jobs to them, etc. at any time and without waiting for the slow validation process characteristic of most blockchains. This does not necessarily mean that everything should be centralised; in fact very little centralisation is required mainly the issues of reconfiguration (adding/removing) and conflict resolution (proof service) and possibly running smart contracts on the blockchain, which we will not discuss here. 

The presence of the infrastructure above the IoT can be captured by introducing into our architecture an actor that we call the Fog Server (FS), which is a local datacenter that has direct connection to \things{} via a suitable radio network (e.g. LoRa), as well as a sufficient compute power, storage and connection to Cloud. The blockchain users should trust the FS for:
\begin{itemize}
\item enrolment of new \things\ to the block chain, and their removal. At the point of enrolment, the Cloud supplies to the FS confidential identity material for the new \thing{} which allows the FS to establish initial credentials of the \thing\ on the blockchain. It also shares with the \thing\ a symmetric key. Also note that a modern IoT device is equipped with hardware encryption facilities and can hold the key in software-unreadable, immutable persistent memory inside its Hardware Security Module (HSM). 

\item as part of the previous, supplying to a \thing\ that joins the blockchain late the authenticated hash of the relevant blockchain history.  

\item withdrawal of a \thing\ from the blockchain. 
\end{itemize}

The FS is the actor that forms blocks for the blockchain by gathering and validating SLVP protocol messages (see section \ref{sec:SLVP}) from \things{} and outside agents (via Cloud) and collecting them into blocks. No trust is required for this, since the FS processes messages received from \things{} and agents on the basis of the blockchain content, which is public and available to all users. Any violation of the validation rules will be noticed by the parties affected and any well equipped {\em witness}, i.e. a (possibly non-enrolled) radio listening post. 

What cannot be protected 100\% reliably by our methods is progress. It is possible in principle for the FS to deny service by refusing to react to valid, legitimate messages sent by \things, or by maliciously modifying the content of those messages, if the FS is compromised. However, server protection is not an issue of IoT security, but a general cybersecurity concern, which is beyond the scope of this paper.

\paragraph{Proxies.} This is yet another zero-trust agent. It has two functions:
\begin{enumerate}
\item Amplify the Sequencer's radio broadcasts (by constructive interference, if LoRa communications are used, or by re-broadcasting broadcast messages). All Proxies are sent the next broadcast message by the Sequencer via the FS just before they are due to be broadcast. 
\item Pick up messages from \things\ directed to the blockchain. This is useful because direct connection to the FS over the air may not be possible given a low power budget and compromise antennae that \things\ have to work with.  A Proxy is connected to the intranet and can forward messages to the server and CAS. The blockchain protocol which we will consider in the next section is robust: Proxies can forward messages of unclear origin and authority; at best they will be filtered out by the FS, at worst they will find their way to the blockchain but will not be properly signed by a legitimate actor and hence will have no effect other than the waste of resources.
\end{enumerate}

All Proxies are connected to a subnet of the local intranet, which facilitates function 1 above as the FS is able to multicast on the subnet provided that the security of the multicast is assured by one of the standard methods that does not involved IoT. 

\paragraph{Communications.} The reader will have noticed that the proposed architecture uses the combination of a dedicated radio channel for security-related data and a general-purpose communication infrastructure, whether wired or wireless, on which \things\ may establish an {\em auxiliary channel}. This is a deliberate choice for the following reasons. 

The purpose of the dedicated radio channel is to ensure by physical means the reliability of broadcast. The only credible threat to the PLS protocol is a DoS attack whereby one or more users of the network cannot get uncorrupted P, L or S messages from the Sequencer. The architecture enables a volume transmission of these messages, with the Sequencer radiating them at maximum legal power and the Proxies joining in by concurrent transmission. We propose that the security radio channel is implemented via LoRa\cite{lora}, a Long Range spread-spectrum technology with many remarkable properties. 

We exploit the fact that as a spread spectrum format, LoRa benefits from concurrent transmissions, where two or more signals carrying the same content under the same modulation regime can be broadcast simultaneously from different locations. Any receiver for whom the signal of one transmitter exceeds the rest only by a factor of 2 (3dB) will not sense the others (\cite{multi-hop-with-CT} p.21436, under ``Capture Effect'', \cite{cotech-interf}). This is due to the Frequency Modulated (chirp) nature of LoRa, and is well known as the {\em capture} effect in both communication and FM broadcast industry \cite{capture}. 

The idea is to endow the Proxies with maximum legal power transmission facilities and to place them in such a way that the communication range of each Proxy captures a certain structural unit on the premises, e.g. a floor (or a building if it is small enough), to form a communication locus. Different loci are separated by distances (as in the case of loci as individual buildings) or obstacles (e.g. construction elements supporting a floor if the loci belong to different floors).  This was studied at length in \cite{multi-hop-with-CT} (see p.21443 under ``The Robustness of CT-LoRa''). The distance is by itself quite an effective dampener, as radio signals fade by 6db when the distance between the transmitter and the receiver is doubled, which would ensure a more than sufficient power contrast for the capture effect.

If properly deployed, the proposed architecture ensures that an attacker can only deny LoRa communications to an IoT node by radiating power far exceeding legal limits, or by installing additional equipment on the premises in violation of physical security. In both cases a large proportion of the IoT devices will remain unaffected by the attack. They will be able to collect all versions of the security messages (including genuine ones and those coming from the attacker) and exchange them between each other using an unprotected network. Security protocols will then quickly establish which versions are genuine.

\section{Posting on the blockchain\label{sec:SLVP}}

\begin{figure}
\hskip-0.5in\bgroup
\def\arraystretch{1.3}%
\begin{tabular}{|c|l|l|l|}
\hline
 \bf Block& \bf Transmit & \bf Verification & \bf BC Action\\
\hline
\hline
$i_0$ &$P_1=H(N_1)$                                    &   Out of Band (Enrolment)                   & \\
\hline
$i_1$& $S_1={\bf E}_{N_1} (H(M_1)\oplus H(N_{2}))$ &               & \\
\hline
$i_2$& $L_1||V_1=H(N_{2}) \oplus N_1\,||\, H(H(N_2)||N_1)$         &                      & \\
\hline

...&...&...&...\\

\hline
\hline
$i_{n}$ &$P_{k}=H(N_{k})$  &...& post $P_{k}$ \\
\hline
$i_{n+1}$& $S_{k}={\bf E}_{N_{k}} (H(M_{k})\oplus H(N_{k+1}))$ &               & \\
\hline
$i_{n+2}$& $L_{k}||V_{k}=H(N_{k+1}) \oplus N_{k}\,||\, H(H(N_{k+1})||N_{k})$         &                      & \\
\hline

\hline
$i_{n+3}$ &$P_{k+1}=H(N_{k+1})$  & Fetch latest $P=P_k$ from block $B=i_n$ & \\
~&~& Set {\bf failed}=true &\\ 
~&~& for $LV$: $B<\#(LV)<i_{n+3}$ &\\ 
~&~& \hskip2em if $H(L\oplus P_{k+1})\ne P$, continue                                       & \\
~&~& \hskip2em set $N = L\oplus P_{k+1}$&\\
~&~& \hskip2em if $H(P_{k+1}||N)=V$       & \\
~&~& \hskip3em {\bf failed}=false; break & \\
~&~& if {\bf failed}, exit & ignore $P_{k+1}$\\
~&~& for $L^\prime V^\prime$: $B<\#(L^\prime V^\prime)<\#(LV)$ &\\
~&~& \hskip2em if $V^\prime=H(L^\prime\oplus N ||N)$, exit & ignore $P_{k+1}$\\
~&~& for all $S$: $B<{\rm \#}(S)<{\rm \#}(LV)$ & \\
~&~& \hskip1em determine  $H_M={\bf D}_N\,S\oplus P_{k+1}$&\\
~&~& \hskip1em send to CAS: $\left(k, {\rm UID}, H_M, \alpha(LV), \alpha(S)\right)$& post $P_{k+1}$\\

~&~& exit & \\
\hline

\hline
\end{tabular}
\egroup
\caption{SLVP Protocol\label{fig:slvp}. Assume bitwise exclusive-or $\oplus$ to have a higher priority than concatenation $||$. $\#(x)$ is the number of the block in which record $x$ is located, and $\alpha(x)$ is its unique ID (address). The sequence of block numbers is in strictly increasing order: $i_0<i_1<i_2<...$ }
\end{figure}

It is tempting to use a protocol similar to what we have described in section \ref{idea} in order for an IoT device to post transactions directly on the blockchain. The advantage of a Guy Fawkes type protocol (which we will call a GF protocol for short) is that any secrets are short-lived, the calculations basic and post-quantum, and communications modest. However the principle vulnerability of such a protocol is its inherent reliance on the precedence of events. With PLS we used wall-clock time to separate intervals; wall-clock time with a reasonable accuracy (less than tens of seconds drift over a year) is cheap and available to even the tiniest of IoT platforms. However, a \thing\ does not generally need to post a transaction on each block of the blockchain, and it is often quite expensive for it to do so from the energy perspective. Precedence can easily be established if the protocol publishes messages on the ledger using an {\em ideal communication environment}, where all messages reach their destination and all can be published in the interval that they were emitted. Which is far from reality in the IoT world. 

\paragraph{Jam-spoof attack.} With Sequencer broadcasting messages for {\em all} users without exception, having maximum legal transmission power and being further supported by Proxies, one can guarantee message delivery (possibly several unauthenticated versions, but that is no problem) either directly, or via subsequent exchanges between users before the interval is out. When a \thing\ sends its content to the FS, the content is of no interest to the peers and the resources used for delivering it are quite limited.  If the \thing\ posts on the chain infrequently, at unpredictable times, any GF protocol based on imperfect communication is potentially vulnerable to the {\em jam-spoof attack}  as follows:

(below $T$ is a \thing\ and $M$ is an attacker)
\begin{enumerate}
\item $T$ runs the protocol to the point where it is about to reveal to a verifier by messaging the so far confidential pre-image to prove a signature (a key feature of any GF protocol)
\item $M$ suppresses the verifier's receiver by jamming the broadcast channel. At the same time $M$ uses a high-gain directional antenna and sophisticated signal reconstruction techniques beyond the capabilities of the verifier to reliably receive the message from $T$. As a result $M$ learns the secret, but the verifier is left believing that the secret has not been revealed yet. 
\item $M$ masquerading as $T$ proceeds to publish its link- and proof-records based on the knowledge of the secret pre-image. The link-message will use a different next nonce than the genuine link-message from $T$ published on the blockchain earlier, but the same current nonce, thus forking $T$'s GF sequence. Also the knowledge of both nonces enables $M$ to post its own signature message on the blockchain to sign an arbitrary hash on behalf of $T$ and to continue to do so indefinitely.
\item $T$ sees the split of its sequence on the blockchain and alerts the FS, but now the FS (or any other arbitrator) is unable, based solely on the blockchain content, to determine whether it is $T$ or $M$ that is the genuine originator of the latest messages. 
\end{enumerate}

Two remedies are available. One is to require authentication of all messages from a \thing\ to the FS. This immediately destroys the zero trust environment we have built, where the only aspect that all users, including the FS itself, have to trust is the correct operation of the Sequencer, and even that up to post-verification. Consequently, we desire to implement a GF protocol that does not require additional trust and which survives a jam-spoof attack.  It turns out that a small modification is sufficient to solve the problem, which brings us to the following SLVP protocol, see fig \ref{fig:slvp}. We will now comment on some of its aspects.

\paragraph{SLVP protocol.} A new blockchain user is {\em enrolled} by the FS\footnote{this makes it a {\em permissioned blockchain}} by authenticating the user's first P-record out of band. Assume that the user sends that record to the blockchain (which means to the FS in the first place) and it appears in some block $i_0$ 

All records on the blockchain carry a UID, i.e. the User IDentification, stated by the message originator. We propose that the user is identified by the first two bytes of its $P_1$ record. The server will not accept the $P_1$ message if it determines that the first two bytes of the hash clash with an already established user. On the other hand, 2 bytes are sufficient for an excess of 64 thousand users, while a typical IoT swarm does not exceed\footnote{There is some evidence that 1000 nodes could saturate the IoT long-range joint channel capacity \cite{lorawan-scale}} 1000. In the sequel we will not distinguish between the UID and the user with a given UID if the context is clear enough to see which we mean.

The table in fig \ref{fig:slvp} presents the protocol from the point of view of a single UID and the FS. Other UIDs will conduct themselves in the same way. Like the PLS protocol introduced earlier the SLVP protocol is invoked periodically, but now at some arbitrary times, which in terms of the blockchain schedule correspond to block intervals. The transmitter does not need a specific confirmation that the message has been received; it simply checks newly formed blocks to find the record that it has transmitted. When this happens, the transmitter sends the next message according to the protocol. If a legitimate transmitter's correct message is not posted on the blockchain after it has been transmitted, it either wasn't received at all, or it was received with an unrecoverable error, or it was impaired in the channel and no longer matches the original message content. Under such circumstances the transmitter will re-send the message until it is posted on the blockchain intact.

The protocol proceeds in rounds, each consisting in three steps: \[
S\to LV\to P\,
\] 
where $S$ is, as before, the signature record, $LV$ is an extended link-verify record, and $P$ is the proof record as before. The $LV$ record consists of the $L$-record, similar to that of the PLS protocol, which is extended with a $V$-record for thwarting the jam-spoof attack. Since the UID is not authenticated and the channel generally lacks integrity, any messages directed to the blockchain can be arbitrarily distorted or lost. 

The SLVP protocol depends on \things' ability to receive blocks on the blockchain successfully via the PLS protocol and any remedial measures that ensure that the block hashes are received and verified before the next Sequencer period by {\em all} users engaging in SLVP. Unlike the Sequencer engaging in timed broadcasts, an SLVP user can be quiescent for many block periods. The new round $k$ starts after the user's $P_k$ message has been received by the FS, validated and posted on the blockchain in some  block $i_n$. Which block this is going to be depends on the  timing of the $P_k$ message. The user, having satisfied itself that its message $P_k$ was received and posted under its UID, sends the $S$-message $S_k$ and waits for it to get posted, too, say in block $i_{n+1}>i_{n}$. The $S$-message does not expose a single bit of either nonce $N_{k,k+1}$ since the value $S_k$ depends on yet undisclosed $H(M_k)$ and since there does not exist an attack on the cipher ${\bf E}$ where neither the key nor the plaintext is known. Sometimes it is convenient for the user to post more than one $S$-record, for example when several documents need signature but they are not otherwise related. The user is allowed to send as many different $S$-messages as is practical. For simplicity, we will assume that one $S$-message is posted in round $k$

Having satisfied itself that the $S_k$ has been posted on the blockchain, the user transmits its $LV$-message. The link part, $L$ is the same as that in PLS, and it serves the same purpose: its value links the current nonce $N_k$ with the new one, $N_{k+1}$. The verify part $V$ is there to make sure that an attacker who learns $N_k$ later cannot combine it with its own $\hat{N}_{k+1}$ and post
\[
\hat{L}_k=H(\hat{N}_{k+1})\oplus N_k
\]
on the blockchain. In such a case the FS would be unable to decide between $L_k$ and $\hat{L}_k$ due to the fact that $L_k$ can be a distorted version of $\hat{L}_k$,  and the message $\hat{L}_k$ an attempt to correct the distortion. With the $V$ message in place, for any pair of $LV$-records: 
\[
L_{k}||V_{k}=H(N_{k+1}) \oplus N_{k}\,||\, H(H(N_{k+1})||N_{k})
\]
and
\[
\hat{L}_{k}||\hat{V}_{k}=H(\hat{N}_{k+1}) \oplus N_{k}\,||\, H(H(\hat{N}_{k+1})||N_{k})\,,
\]
where 
\[
H(L_k\oplus P_{k+1}) = H(\hat{L}_k\oplus P_{k+1}) = P_k\,, 
\]
the one posted in an earlier block wins: the protocol-compliant user does not disclose the genuine $P_{k+1}=H(N_{k+1})$ in the same block as $L_{k}||V_{k}$ 
and so the fact that \[
V_k=H( P_{k+1}|| L_k\oplus P_{k+1})
\]
proves that the originator knew $H(N_{k+1})$ before it was posted. The only actor that knows $H(N_{k+1})$ before it is posted is the genuine user.

The message $P_{k+1}$ is sent in some later block period $i_{n+3}$ to finish the current round. The FS verifies the message and processes the round that it completes along the lines of the above argument. If verification succeeds, the new $P$-record is posted. Notice that no matter how many counterfeit $LV$ messages have been posted by attackers since the last verified $P$ and no matter how many counterfeit $P$-records are sent after them, only one $P$ record will be accepted and posted by the FS in any round of the protocol on behalf of any given UID. Also the FS will find only one $LV$-record to be valid, which is the earliest $LV$-record compatible with both the previous and the newly validated $P$-record. 

Next the FS will reverse all $S$-records posted after block $i_n$ under the user's UID  by computing \[
H_M=P_{k+1}\oplus {\bf D}_{P_{k+1}\oplus L}S\,. 
\]

Each record $r$ on the blockchain has its address $\alpha(r)=(i_r,l_r)$, where $i_r$ is the block number in which $r$ is located and $l_r$ is the sequential number of $r$ among the records of the same type and under the same UID in block $i_r$. 

For each record $S$ in the current round $k$, the FS collects proof data in the following form: 
\[
W_S = (k,{\rm UID},H_M,\alpha(LV),\alpha(S))  
\]
and instructs the CAS unit to store $W_S$ under $H(W_S)$ as usual. The CAS unit will use $H_M$ as a {\em trigger}. When/if the user UID stores content $c$ in CAS, such that $H(c)=H_M$, the CAS manager will post a special $C$-record, on the blockchain on behalf of UID as follows:
\[
C = {\rm UID}:(H_M,H(W_S))   
\]
which serves as blockchain confirmation that CAS has taken charge of the content file as well as the proof data for it for any witness to verify. Triggers that are not triggered by the user over a certain number of blocks (large enough to conclude that the original $S$-message was counterfeit/distorted) are removed from CAS and copied to the FS security log. A $C$-message will be ignored (and the corresponding $C$-record not posted) if a trigger for it has not been provided by the FS at the time of submission. 

Counterfeit $S$-messages pose no threat. They will be deciphered to an unpredictable $H_M$, and an attacker would not be able to provide content that matches a given hash value anymore than it is able to find a nonce $N$ given $H(N)$, the latter being the main security assumption for any GF protocol. 

\section{Enrolment and optimisations\label{enrol}}

It has been mentioned earlier that the very first hash $P_1$ of the protocols is validated out of band. For a user to be able to start SLVP there are two requirements: 
\begin{itemize}
\item access to the blockchain which includes out of band validation of the latest $P_k$ and all previous blocks from 1 to $k-1$, inclusively
\item registration of the user's $P_1$ for out of band validation.
\end{itemize}

Enrolment of new equipment normally requires a human administrator as it involves physical placement, configuration and initialisation of the item  according to the business objectives. We propose the following enrolment protocol:
\begin{enumerate}
\item The administrator's workstation establishes secure confidential communication with the FS using Cloud and state of the art security. The FS shares a fresh key $K$ with the administrator.
\item The new \thing\ that the administrator has ascertained to be genuine 
\begin{enumerate}[i]
\item receives $K$ and the Sequencer's latest $P_k$ using near-field communications (NFC) or similar, 
\item generates $N_1$ and another random nonce $N^*$ 
\item computes $P_1=H(N_1)$ 
\item sends $Q=P_1||{\bf E}_K (P_1 \oplus N^*)$ back to the server via the administrator's NFC port acting as a relay. 
\end{enumerate}

\item The FS examines a short prefix of $P_1$, $\pi(P_1)$, e.g. 2 bytes, and checks that no UID with this value has been enrolled. If that is the case, the server computes $N^*$ from Q and responds with ${\rm ACK}=H(N^*)$ otherwise the response is FAIL. 

\item If the response is FAIL, the \thing\ generates a new pair $N_1$ and $N^*$ and repeats steps 2 and 3.
Otherwise
\begin{enumerate}[i]
\item the \thing\ verifies that ${\rm ACK}=H(N^*)$ and notes its new UID, i.e. $\pi(P_1)$
\item confirms completion to the administrator.
\end{enumerate}
If ${\rm ACK}\ne H(N^*)$ the protocol fails; a notification to this effect quoting $P_1$,$N^*$ and ACK is sent to the administrator for subsequent analysis.

\end{enumerate}

Now the new IoT device is ready to receive the minimum data necessary to access the blockchain. The amount of trust required for it is exactly the same as it is for any other user: it needs to authenticate the latest $P_k$, except for the devices that have been present from the start the index $k=1$. But how is it going to authenticate the blocks that were formed {\em before} block $k$? 

\begin{figure}
\begin{center}
\includegraphics[width=0.8\textwidth]{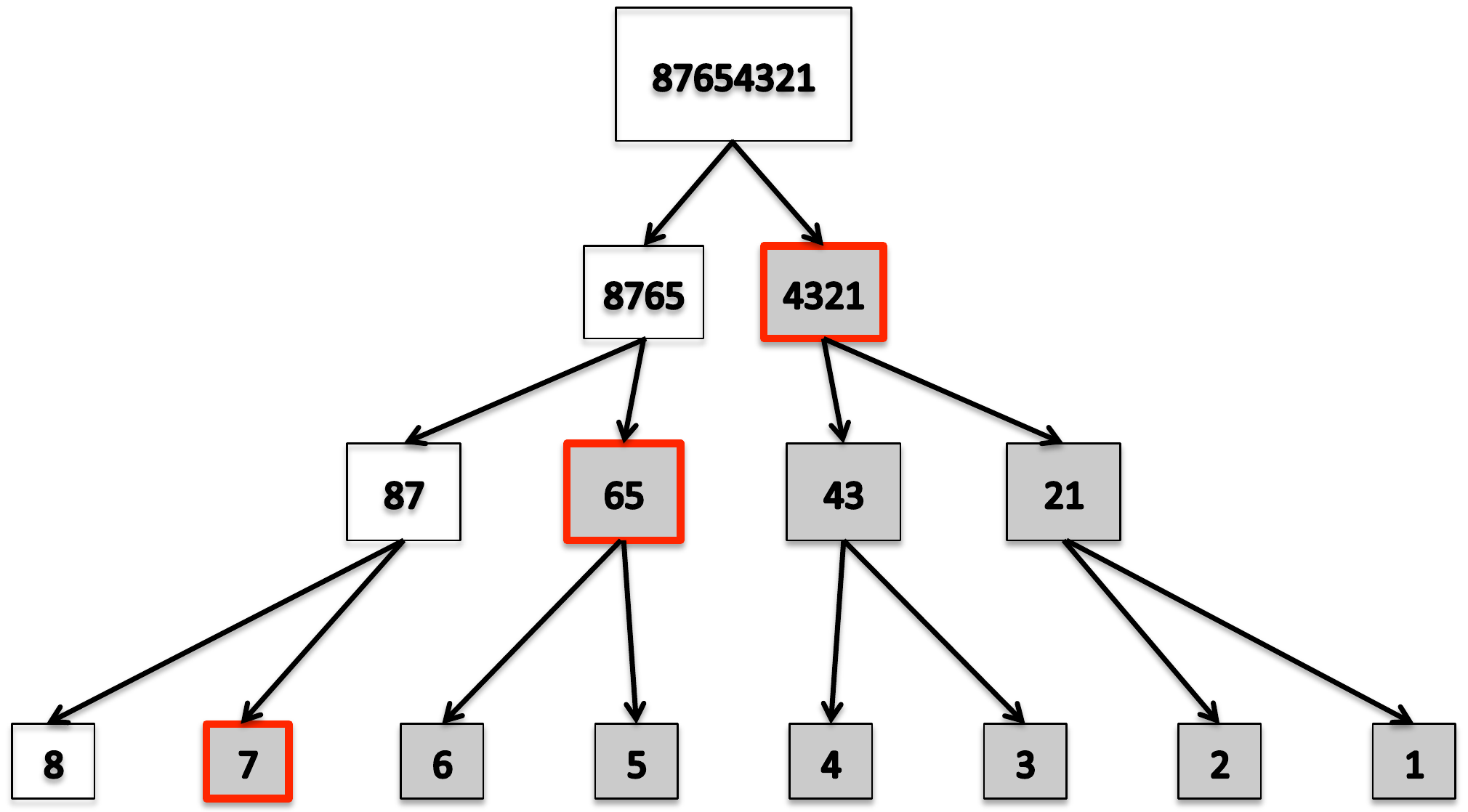}
\end{center}
\caption{Merkle forest\label{fig:merkel}}
\end{figure}

We propose to communicate the Merkle minimal forest roots of the current state in each blockchain block. The roots are a compact collection of root hashes that can be followed on CAS to securely access any block from $B_1$ to current stored in CAS. To illustrate the concept, let us imagine a block sequence from block 1 (initial) to block 7 (current), see fig \ref{fig:merkel}. For simplicity we  use a binary Merkle tree, in which every node is composed of the hashes of the two child nodes' contents. Some of the nodes are already formed and will never change (shaded in the figure), and some are still being formed pending the future blocks. It is easy to see that the Merkle proof of any block up to and including 7 requires only hashes of node 4321, 65, and 7 as roots, the paths of every leaf from 1 to 7 is rooted at one of them. Notice that the binary representation of 7 is 111 which corresponds to one node each at levels 0,1 and 2. For block 5=$101_2$ we would have a level-2 block and a level-0 block, which agrees with the diagram. 
  
A binary Merkle tree is unjustifiably deep. Focusing on the world of the IoT, we recognise that communications are typically limited to messages no longer than 200-250 bytes, so given a typical hash size of 32 bytes, it is convenient to use a quad-tree, which will be much less deep. For a quad tree in a blockchain of say, 1 mln blocks (at one block per 15 min, this gives us more than 10 years' running), we get $\log_4 10^6\approx 10$, which means that the server only needs to authenticate at most 10 hashes to give a new IoT \thing\ a secure start. For every $k$ the record consisting of the minimal root set 
\[
\Gamma(k)=(r_1,...,r_p)
\]
is computed and stored by the FS under $\gamma_k=H(\Gamma(k))$ in CAS, where $k$ is the current block number, and $r_1,\ldots, r_p$ are the Merkle tree hashes that correspond to the nonzero digits in the base-4 representation of the number $k$. 

For each $k$, the FS will  put $\gamma_k$ at the beginning of block $k+1$. Any user that wishes to trade storage for CAS communication, or a new user who has missed an initial segment of the block chain, but who trusts the latest $P_k$ can use the $\gamma$-record on the block to securely retrieve any preceding block(s) via CAS, {\bf if} they choose to trust the FS. We would like to remark that the FS in this particular case {\bf is} trustworthy, since the $\gamma$ record can be computed by any full witness of the blockchain (i.e. any user that has been present since block 1) and if the FS is compromised, the proof of that will be constructed instantly. 

Regarding the storage requirement in CAS, they are minimal. There is no duplication due to the very nature of CAS. Summing up the geometric series for the radix-4 Merkle tree with depth 10, we get circa 350,000 hashes to store for $10^6$ blocks, about 10Mb, a trivial amount of storage.

\subsection{Countermeasures against DoS} 

The acquisition of a shared secret between the \thing\ and the FS at the point of enrolment does not make the blockchain any less useful. Indeed, in our threat model the FS is not trusted by any enrolled user any more than any other user of the blockchain, so the shared secret cannot be used to replace the security protocols that make the blocks of the blockchain an immutable, ordered, authenticated sequence of records. Nor is it any good for non-repudiation. However, just as the Sequencer is trusted to keep its secret for the avoidance of blockchain split so is the FS trusted to be interested in reducing the amount of {\em noise} on the blockchain, i.e. records sent in by an attacker on behalf of a genuine UID, which will eventually be caught out and eliminated by the SLVP protocol. After all, as the FS is solely responsible for what does and what does not get posted, the proposed blockchain concept only works on the assumption that the FS itself is {\em not} and can never be behind a DoS attack. The assumption that the FS will be a willing party to an additional noise-reduction protocol does not add much to that.   

With this in mind we propose that each \thing\ uses a very short Message Authentication Code (MAC) based on symmetric encryption and the shared  key $K$ received at enrolment. The MAC need not be longer than 2 bytes (possibly even 1 byte) and can be computed using standard techniques by the \thing's hardware security module or crypto accelerator. The MAC is computed for each message of the SLVP protocol sent to the server as well as the content messages sent via the FS to CAS. Due to the shortness of the MAC, the exposure of the shared key is minimal, obviating session keys. If the MAC does not match, the FS ignores the message. With a 2-byte MAC, an attacker would have to send tens of thousands of messages to get through to the FS in the first instance; such a volume on behalf of a single IoT user will surely raise the alarm, resulting in the intruder's triangulation and suppression.

Recall that the Sequencer's messages may arrive distorted or not arrive at all, and the users, especially \things, must talk to each other to collect a set of versions for each PLS message to ensure that the set contains the original. To facilitate this, a short authenticator can be sent by a Proxy on an auxiliary channel to each \thing\  by transmitting \[
u=\hbox{UID},\hbox{\bf cat},\pi(H(M))
\]
where UID is its User ID, $\pi(H(M))$ is a short hash of message $M$ from the category {\bf cat} (one of $P$, $L$, or $S$). Message $u$ is extended with $\hbox{MAC}_K(u)$, where $K$ is the  key agreed with UID at enrolment. The message $u$ is prepared by the server and is forwarded by one or more of the Proxies on the auxiliary channel. User UID, having received $u$ and checked the MAC, recalculates $\pi(H(M))$ based on the latest message in category {\bf cat} received (if it did at all) and compares it with the value contained in $u$. If they match, the device joins a re-broadcast concurrent-transmission group on a pre-arranged channel (frequency and time relative to the start of the Sequencer broadcast interval) to help nearby nodes with PLS reception. Given that PLS messages are short (not much longer than 32 bytes if SHA-256 is used for $H(\cdot)$) and infrequent (3 messages typically 2--5 times per hour, $~0.5$ KB/hour), a blockchain supported \thing\ can afford to transmit as much to help other \things\ (which in turn will help it) to survive a DoS attack.

When it comes to the SLVP protocol, the user is the active transmitter, and the roles are reversed. Now as a DoS resilience measure, the user UID adds a $\hbox{MAC}_K(x)$ to every message $x$ that it sends to the server (possibly via a Proxy) for posting on the blockchain. The FS checks the MAC based on the received UID and the shared key $K$ and if the MAC does not match, it ignores $x$. Again, we must stress that if the MAC does match, this means nothing in terms of the SLVP protocol, since the FS does not trust the \thing\ any more than the \thing\ trusts the FS. Reduction of noise is their common concern: the FS acts on behalf of the owner of the IoT network and is interested in suppression of a DoS attacker, and the \thing\ will keep its shared $K$ secret to avoid an attacker's spoofing it and preventing its legitimate messages from reaching the blockchain. Commonality of concern is the only reason why the additional authentication will be effective.

\section{Emergency mode\label{emode}}

A distinguishing feature of IoT is its multiplicity of time scales. Most \things\ require only infrequent interaction with the outside world, reporting sensor readings, receiving parameter updates and possibly code upgrades. All these activities are easily accommodated by the blockchain mechanism and are protected by its inherent security properties. A major downside of a blockchain is its latency. No matter how frequently new blocks are added to the chain (and in our case they are not even mined), a \thing\ may discover itself in a situation when it must raise the alarm  sooner than a new block can be published, especially since in the case of the PLS blockchain, blocks are published on a fixed wall-clock schedule. Even if a new block is to emerge soon, there is no guarantee that any given \thing\ will be able to post its message in it rather than a later block. 

This problem is quite practical: a hospital monitor detecting a catastrophic change in a patient's condition and a nuclear plant's sensor detecting a reactor malfunctioning are cases in point to name but two. We emphasise that emergency messages are {\em not} an alternative of posting records on the blockchain. The latter is more powerful in that \things\ are able to securely interact with each other directly via their signed blockchain messages, whereas emergency communications are targeted solely at the FS for consumption outside the IoT network.

One might think that emergency communications can be supported by the shared key $K$ that the originating \thing\ agreed at enrolment. Indeed the FS can request a full MAC and satisfy itself that the message is authentic. However, this is not enough. Emergency communications involve rapid response and that can only be provided if an independent arbitrator can establish that the message was sent by {\em no-one but} the claimed originator. In other words, a signature rather than mere authentication is required. In the absence of signature, the response agent would be running the risk of the originator repudiating the message: after all, the symmetric key $K$ is shared with the FS, and so either the FS or an agent to which the FS has leaked the key (willingly or not) might have sent the emergency message.

Non-repudiation is not a concern with blockchain communications, they cannot be repudiated for obvious reasons. However, post hoc validation by blockchain is only useful for confirmation of valid messages, rather than proving a message to be invalid, since the rapid response must come into effect {\em before} blockchain validation may take place. The other way\footnote{Our design constraint 1, Post Quantum, prevents standard public-key cryptography, which would provide an effective signature if the originator's public key is validated in advance by blockchain} of ensuring non-repudiation is by One–Time Signature (OTS), which we will consider next.

\subsection{OTS}

OTSs are known to have a very large ``public key'', i.e. authenticated public data used for validation of a signature. In the original OTS proposed by Lamport \cite{OTS}, the originator shares with the verifier $k$ pairs \[
(H(n_1),H(N_1)), (H(n_2),H(N_2)),\ldots, (H(n_k),H(N_k))\,,
\] 
where all $n_i$, $N_i$ are random nonces. To sign a $k$-bit message $\{x_i\}$, the originator additionally supplies $k$ values $\{s_i\}$:
\[
s_i = 
\begin{dcases}
n_i& \hbox{if}\; x_i=0\\ 
N_i& \hbox{otherwise} 
\end{dcases}
\] 

OTSs solve the problem of emergency non-repudiation if the public key is signed and posted on the blockchain in advance (using SLVP), but the price for an IoT device using it is prohibitive. A straightforward application of OTS to signing a full hash of an emergency message would require $256\times256\times 2=128$K bits of public key, or 16KB. It is easy to see that the public key can only be used once if we want the security of hash pre-image to work for every bit of a signed message. Even if using the large key once were acceptable (think of a catastrophic circumstances that do not present themselves often), the {\em signature }size of would still be half as large: 8KB, which would take some time to communicate over a low bit-rate channel, especially in the presence of transmission errors necessitating a re-transmission. 

\subsection{Public key}
Let us start with the public key problem. We propose to bring the SLVP protocol messages to bear on the emergency mode to eliminate transmission and authentication of the public key. Recall that a \thing\ running the protocol sends messages that depend on nonces $N_k$ which are chosen by it at random, see figure \ref{fig:slvp}. At the validation step, the FS computes $\hat{N}_k=L_k\oplus P_{k+1}$ and verifies that $\hat{N}_k=N_k$ by applying $H(\cdot)$ to both sides and checking the equality. 

We propose that every \thing\ engaging in SLVP must compute random nonces $N_k$ by building a hash chain:
\begin{align*}
N_k&=N_k^{[\alpha]}\\
N_k^{[i]}&=H(N_k^{[i-1]})\, \hbox{where}\; i=1,\ldots,\alpha\;,
\end{align*}
and where $N_k^{[0]}$ is completely random and is kept secret by the \thing\ for at least $\alpha$ rounds of the protocol.

In other words, every nonce is an image of a random number under $\alpha$ applications of $H(\cdot)$, which is known as the Winternitz chain. When the server has posted the value $P_k$ it received from the \thing, it has access to, and has verified, 
\[
N^{[\alpha]}_i\, \hbox{for all}\; i=0,\ldots,k-1
\]

For a given UID, the private key for the period between the postings of $P_k$ and $P_{k+1}$ (i.e. when nonces up to and including $N_{k-1}$ have been revealed) is as follows:
\[
\{N_{k-i}^{[\alpha-i-1]}\}\; \hbox{for}\; i=1,\ldots,\alpha-1\,.
\]  

\begin{figure}
\begin{center}
\includegraphics[width=0.4\textwidth]{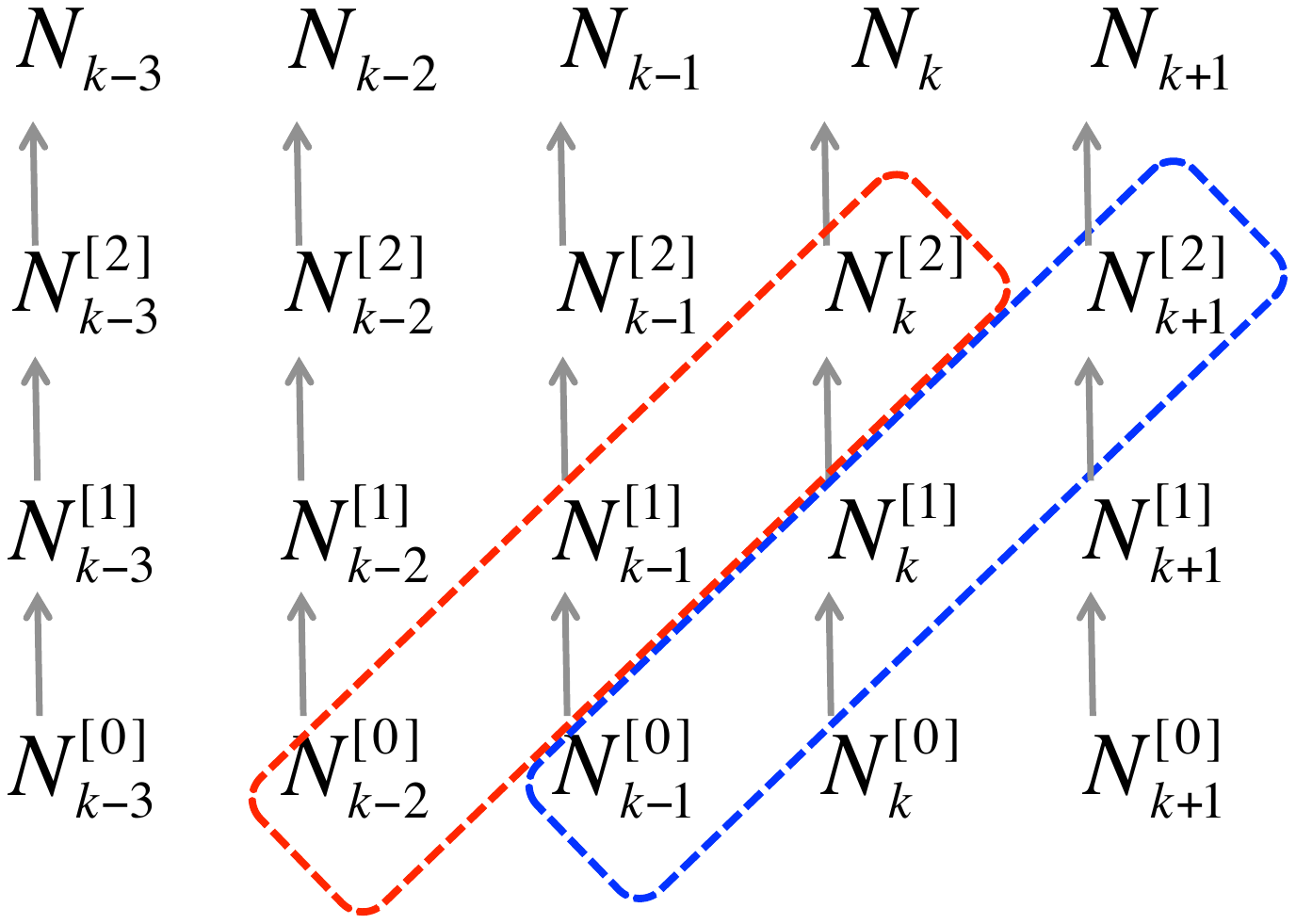}
\caption{Use of Winternitz chains for OTS private key production from SLVP nonces at round $k$ (red) and $k+1$ (blue) for $\alpha=3$. NB: $N^{[3]}_i=N_i$ so the top row is shown without superscripts. Vertical arrows signify application of $H(\cdot)$.
\label{fig:hors}}
\end{center}
\end{figure}

If Lamport's OTS is used, the \thing\ sends to the FS a selection of the values  $N_{k-i}^{[\alpha-i-1]}$. The FS verifies each value by applying $H(\cdot)$ to it $i+1$ times and comparing the result with $N_{k-i}^{[\alpha]}=N_{k-i}$ that it has at its disposal for the same UID. Notice that as $k$ advances to $k+1$ with a new round of SLVP, the same chain $N_{k-i}$ is used for public key with an earlier pre-image:
\[
N_{k-i}^{[\alpha-i]}\,\,\to\,N_{(k+1)-(i+1)}^{[\alpha-(i+1)]}\,
\]
until the protocol is $\alpha$ rounds ahead of the chain at which point the chain will have been fully used and is no longer required for OTS purposes. An example of $\alpha=3$ is displayed in figure \ref{fig:hors}, showing two consecutive rounds. Notice that the values used in a later round are always lower on their Winternitz chain than those in earlier rounds, making them secure within the second pre-image hardness assumption.
We conclude that the ``private key'' for an SLVP round, i.e. the set of potential pre-images to be used for OTS, is unknown to the FS in that round and that the FS has access to the authenticated public key to verify the signature. 

This arrangement of chains and pre-images makes it possible for a \thing\ running SLVP {\em not to share any public key at all} and at the same time be able to sign messages in emergency mode without waiting for blocks to appear on the blockchain.  It is quite useful for the IoT world, and the price that we pay is the need to pre-hash a random string $\alpha$ times at every round of the SLVP protocol rather than use it directly as a nonce. Taking a popular ESP32 system-on-cheap as a specific example we learn from \cite{esp32} that it takes $\sim1\mu$s at full power to process one AES256 hash block, perhaps 0.1ms for $\alpha=100$. The energy spent is a fraction of LoRa communication cost for the same. It is completely justified if the \thing\ potentially requires emergency communications in this or any of the future $\alpha$ rounds, assuming that it is sufficient to sign a certain number of bits $L$ of the emergency message (or its digest) to reassure the responder of nonrepudiation. For Lamport's classical OTS signature, $L=\alpha/2$. We will improve on this next.

\subsection{GF-HORS}

The excessive size of OTS signatures have been recognised by many authors, and several proposals have been made to improve on this. We will follow the methodology presented in \cite{HORS}, where an original idea, Hash to Obtain a Random Subset (HORS), was first put forward.

Assume that $\alpha$ is a power of 2. Compute a length-$L$ digest of the message we wish to sign, and partition its binary representation into slices $\log_2 \alpha$ bits long. Interpret these slices as unsigned numbers 
\[
\sigma_j,\; j=0,\ldots,{L\over \log_2 \alpha}-1\,.
\]
where all $\sigma_j<\alpha$.
Now for each  $j$ the \thing\ supplies the value  $N_{k-\sigma_j}^{[\alpha-\sigma_j-1]}$ to form a signature. The FS validates the signature by recomputing the digest of the message, then recomputing $\{\sigma_j\}$ from the digest, and then for each $j$ verifying $N_{k-\sigma_j}^{[\alpha-\sigma_j-1]}$ by applying $H(\cdot)$ to it $\sigma_j+1$ times and checking that the result equals the previously obtained nonce $N_{k-\sigma_j}$.

The idea to use the digest of the message to be signed rather than the actual bits of it by partitioning the string was first proposed in \cite{HORS}, and the security of this method is slightly less than that of the second pre-image hardness, since here the attacker only needs to find a message whose digest partitioned into suitable chunks gives the same {\em set} or even a subset of $\{\sigma_j\}$ in any order. However, the authors of \cite{HORS} remark that finding a (useful) message that has the same set or a subset of digest chunks as a given one is still computationally hard for a good digest. 

We propose a GF-HORS signature (HORS signature with SLVP-derived public key) based on a {\em keyed MAC with the shared key $K$} as the digest. The MAC protects the message being signed from an outside forgery, and the HORS signature protects it from an insider job. Let us take a look at some example numbers to illustrate the efficiency of the scheme.

If we assume $\alpha=64$ and use AES-128 for the digest MAC (shortening it down to 126), we get up to $21$ sigmas. Assuming for estimation purposes that the digest is a random bit string, the probability that an attacker's digest gives a subset of the sigmas, is less than $(21/64)^{21}$, around $10^{-10}$, a pretty good result for an IoT device nonrepudiation. The communication cost of the signature is $256\times21/8=672$ bytes, about three messages on LoRa. Recall that we require three shorter messages (around 128 bytes all together) for an SLVP round, which is in the same order of magnitude.

A final remark. When a \thing\ is first enrolled by the server, there is not enough nonces in its history (in fact there aren't any initially) for the formation of the public key. One remedy could be to produce $\alpha$ nonce chains at enrolment and share $\alpha$ chain-ends with the FS at that point. Another solution is to consider the first $\alpha$ SLVP rounds of a new \thing\ a probationary period, when it is being tested and adapted to its environment and when it is not allowed to participate in emergency communications.

\section{Related work\label{related-work}}

The advantages of blockchain technology in the case of IoT are not clearly articulated in literature. Recent surveys \cite{IoT-Sec2018},\cite{IoT-Blockchain2019} recognise blockchain as a disruptive technology for the IoT, and list the benefits in generic terms:

\begin{itemize}
\item Decentralisation: Distributed Ledger Technology is supposed to be more robust and secure against a single point of failure.
\item Pseudonymity: the ability to enrol a new actor by registering its public key (or public hash, in our case)
\item Security of Transactions. This boils down to the immutability of the ledger.
\end{itemize}

This is matched with a plethora of use cases mentioned in \cite{IoT-Sec2018}, see pp. 212-214. However, none of the bullet points is specific for the IoT.

We find our objectives to be close to those of \cite{smart-home}, and that paper is a good illustration of how different our approach is from the direction inspired by the typical assumptions. The authors of \cite{smart-home} assume, like others (see, for example, \cite{lightweight-clients}), that an individual IoT device is likely to be underpowered for managing blockchain transactions {\em directly}, as it does not have the storage space, communication bandwidth or processing power for such a task. 

As far as communications are concerned, article \cite{smart-home} correctly posits that low bit-rate radio channels, such as LoRa will be used. However it pays to differentiate between communication of a small amount of security-related data and unsecured, bulk public data transfer.

Storage-wise, to the best of our knowledge published research assumes that the blockchain either has to be stored at the IoT device itself (which is indeed expensive), or else trust must exist between the device and any storage server. 
The latter assumption is not necessarily justified due to the availability of Content-Addressable Storage (CAS), which is, by construction, self-certified not requiring trust or secure communications. The idea of CAS goes back to the late 1990's paper \cite{first-cas-crc} where it was proposed to use a file's CRC as its name, which is not quite satisfactory due to massive aliasing, but a few years later paper \cite{first-cas-crypto} suggested the cryptographic hashes of files should serve as file names. In the last five years the leading general-purpose CAS project has been one known as InterPlanetary File System (IPFS) \cite{IPFS} and it is widely used. 

The original Guy-Fawkes protocol on which PLS is based (\cite{GFP}, p.12) requires four items to be published in every round of the protocol, while PLS only publishes three. Also, verification in a round of Guy Fawkes requires a calculation that involves three items to be hashed together, whereas PLS computes a hash of one item of a minimum size, a factor of three saving on the receive side. PLS performs a symmetric decryption to obtain and confirm the message (or, to be precise, the message hash), which Guy Fawkes does not have to do. However, taking an example of ESP32 \cite{esp32} as a popular system-on-chip for IoT with a crypto accelerator, the AES-256 decryption calculation costs at most 22 clock cycles, while computing SHA-256 requires at least 60 clock cycles to process one block plus a minimum of 8 cycles to produce the digest. 
This means that a hash is at least three times as expensive as the standard encryption. This is to do with the nature of the algorithms (a much smaller number of rounds for encryption compared to hashing), and it is likely that a hash is several times slower than a symmetric encryption on other architectures as well. 

We conclude that PLS is both faster and less communication-intensive at the receiver end. At the transmitter end performance matters little, since the FS and Sequencer are not on a tight energy budget. 

We are aware of one prior attempt at using a GF protocol in conjunction with a blockchain: \cite{Fawkscoin}. In that paper the blockchain itself is assumed to be Bitcoin and a GF protocol is used only for signing value transfer messages (i.e. transactions). The authors of \cite{Fawkscoin} were aware of the jam-spoof attack (which they call {\em race-condition theft}), but their solution is partial, based on a time-out whereas the $V$-messages in our SLVP protocol capture both pre-images, the current and the next ones, defeating the jam-spoof attack without needing a time-out facility (but we still require the ``earlier LV message wins'' analysis similar to \cite{Fawkscoin}).   

Repeated application of the hash function in order to reduce the size of the public key was first suggested by Winternitz according to Merkle\cite{hashsig}. We are not aware of any prior work on our proposed sliding window across Winternitz chains. We use the original HORS\cite{HORS} procedure, but this has been improved to HORST \cite{SPHINCS}, which we can also accommodate. HORST differs from HORS by the fact that the public key is stored in a Merkle tree, whose root is authenticated in advance (at the cost of one round of the SLVP protocol in our case). We prefer the original, HORS, as it allows us to piggy-back the public key on the SLVP nonce sequence by producing each nonce off the top of an individual Winternitz chain with the bottom kept confidential; as a result the user does not need to publish its public key at all. Not only does it save us a round of SLVP, it obviates communication of a large public key to CAS as well. However, we recognise that HORST may be useful for emergency messages if the required long-term security necessitates a much longer signature, in which case the user must store a sufficiently large HORST Merkle tree in CAS in advance and bear the risk of exposing the private key, held inside an IoT device, to a physical intruder.   

\section*{Conclusions}

We have presented the architecture and protocol suite for a permissioned blockchain construction based on the Guy Fawkes family of protocols. Our construction requires limited trust for one sealed, air-gapped unit we call Sequencer which is not internet-connected and which is responsible for keeping a short-term secret. If the short-term secret is kept, we show that this type of blockchain will not split and will maintain immutability. The rest of the network is untrusted. 

We have proposed a protocol for  posting signed messages of any size on the blockchain without using public-key cryptography and discussed its security. Finally, we have shown how emergency (zero-latency) communications can coexist with a PLS blockchain without requiring a public key, yet maintaining nonrepudiation.

The main threat to a PLS blockchain is DoS attacks. While those cannot be fully eliminated for a radio network susceptible to jamming, we suggested the use of a shared secret for non-physical DoS defence: reduction of the number of counterfeit regular and emergency messages accepted for analysis. The outcome of that analysis does not depend on the security of the shared secret, but the efficiency does. Our threat model assumes that sustained physical and non-physical attacks will trigger direction-finding and triangulation of the signal source, eventually eliminating the threat. The advantages of our proposed method are the following:
\begin{enumerate}
\item GF protocols have the convenience of public-key crypto without having to manage keys, perform costly computations on underpowered IoT devices or be vulnerable to quantum attacks. 
\item \things{} can validate each other's transactions without trusting third parties. 
\item even though the blocks are not mined, passive receivers of the authenticated block content (which we call witnesses) that monitor radio communications are able to detect wrongly accepted records and raise an alarm without the owner (Fog Server) being aware of the monitoring process.
\item validation proofs become objects in their own right, stored in the same CAS structure as all other blockchain records; they can be confirmed by any witness of the blockchain at low cost based solely on their content. 
\end{enumerate}

\iftrue
Future work will define mechanisms and protocols for managing trust whereby a \thing{} may delegate verification of 
transactions to a blockchain witness. We will also focus on a lightweight smart-contract language.

The author is grateful to Professor Bruce Christianson for useful discussions and his comments on the manuscript.
\fi

\bibliographystyle{plain}
\bibliography{GF-BC}
\end{document}